\documentclass[twoside,twocolumn,9pt]{article}
\usepackage{extsizes}
\usepackage[super,sort&compress,comma]{natbib} 
\usepackage[version=3]{mhchem}
\usepackage[left=1.5cm, right=1.5cm, top=1.785cm, bottom=2.0cm]{geometry}
\usepackage{balance}
\usepackage{times,mathptmx}
\usepackage{sectsty}
\usepackage{graphicx} 
\usepackage{lastpage}
\usepackage[format=plain,justification=justified,singlelinecheck=false,font={stretch=1.125,small,sf},labelfont=bf,labelsep=space]{caption}
\usepackage{float}
\usepackage{fancyhdr}
\usepackage{fnpos}
\usepackage[english]{babel}
\addto{\captionsenglish}{%
  \renewcommand{\refname}{Notes and references}
}
\usepackage{array}
\usepackage{droidsans}
\usepackage{charter}
\usepackage[T1]{fontenc}
\usepackage[usenames,dvipsnames]{xcolor}
\usepackage{setspace}
\usepackage[compact]{titlesec}
\usepackage{hyperref}
\usepackage{amsmath}
\usepackage{amssymb}
\usepackage[dvipsnames]{xcolor}

\usepackage{epstopdf}

\definecolor{cream}{RGB}{222,217,201}

\begin{document}

\pagestyle{fancy}
\thispagestyle{plain}
\fancypagestyle{plain}{

\fancyhead[C]{\includegraphics[width=18.5cm]{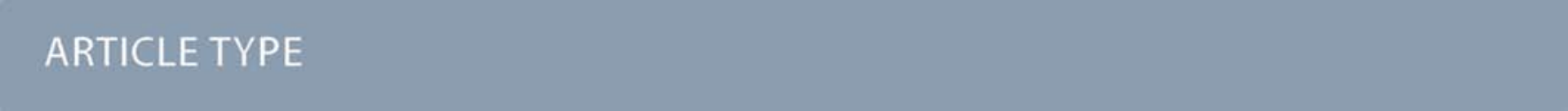}}
\fancyhead[L]{\hspace{0cm}\vspace{1.5cm}\includegraphics[height=30pt]{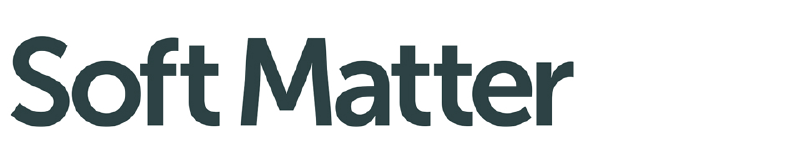}}
\fancyhead[R]{\hspace{0cm}\vspace{1.7cm}\includegraphics[height=55pt]{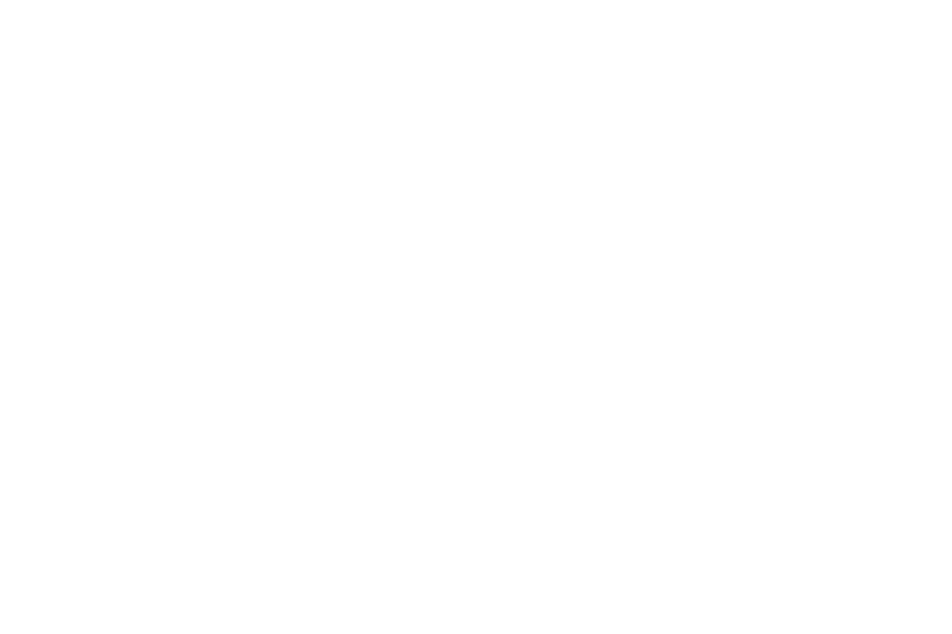}}
\renewcommand{\headrulewidth}{0pt}
}

\makeFNbottom
\makeatletter
\renewcommand\LARGE{\@setfontsize\LARGE{15pt}{17}}
\renewcommand\Large{\@setfontsize\Large{12pt}{14}}
\renewcommand\large{\@setfontsize\large{10pt}{12}}
\renewcommand\footnotesize{\@setfontsize\footnotesize{7pt}{10}}
\renewcommand\scriptsize{\@setfontsize\scriptsize{7pt}{7}}
\makeatother

\renewcommand{\thefootnote}{\fnsymbol{footnote}}
\renewcommand\footnoterule{\vspace*{1pt}%
\color{cream}\hrule width 3.5in height 0.4pt \color{black} \vspace*{5pt}} 
\setcounter{secnumdepth}{5}

\makeatletter 
\renewcommand\@biblabel[1]{#1}            
\renewcommand\@makefntext[1]%
{\noindent\makebox[0pt][r]{\@thefnmark\,}#1}
\makeatother 
\renewcommand{\figurename}{\small{Fig.}~}
\sectionfont{\sffamily\Large}
\subsectionfont{\normalsize}
\subsubsectionfont{\bf}
\setstretch{1.125} 
\setlength{\skip\footins}{0.8cm}
\setlength{\footnotesep}{0.25cm}
\setlength{\jot}{10pt}
\titlespacing*{\section}{0pt}{4pt}{4pt}
\titlespacing*{\subsection}{0pt}{15pt}{1pt}

\fancyfoot{}
\fancyfoot[LO,RE]{\vspace{-7.1pt}\includegraphics[height=9pt]{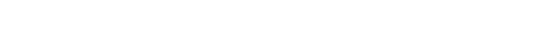}}
\fancyfoot[CO]{\vspace{-7.1pt}\hspace{13.2cm}\includegraphics{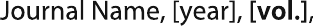}}
\fancyfoot[CE]{\vspace{-7.2pt}\hspace{-14.2cm}\includegraphics{RF}}
\fancyfoot[RO]{\footnotesize{\sffamily{1--\pageref{LastPage} ~\textbar  \hspace{2pt}\thepage}}}
\fancyfoot[LE]{\footnotesize{\sffamily{\thepage~\textbar\hspace{3.45cm} 1--\pageref{LastPage}}}}
\fancyhead{}
\renewcommand{\headrulewidth}{0pt} 
\renewcommand{\footrulewidth}{0pt}
\setlength{\arrayrulewidth}{1pt}
\setlength{\columnsep}{6.5mm}
\setlength\bibsep{1pt}

\makeatletter 
\newlength{\figrulesep} 
\setlength{\figrulesep}{0.5\textfloatsep} 

\newcommand{\topfigrule}{\vspace*{-1pt}%
\noindent{\color{cream}\rule[-\figrulesep]{\columnwidth}{1.5pt}} }

\newcommand{\botfigrule}{\vspace*{-2pt}%
\noindent{\color{cream}\rule[\figrulesep]{\columnwidth}{1.5pt}} }

\newcommand{\dblfigrule}{\vspace*{-1pt}%
\noindent{\color{cream}\rule[-\figrulesep]{\textwidth}{1.5pt}} }

\makeatother

\twocolumn[
  \begin{@twocolumnfalse}
\vspace{3cm}
\sffamily
\begin{tabular}{m{4.5cm} p{13.5cm} }

\includegraphics{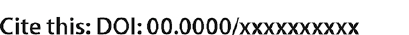} & \noindent\LARGE{\textbf{The role of deformability in determining the structural and mechanical properties of bubbles and emulsions}} \\
 & \vspace{0.3cm} \\

	 & \noindent\large{Arman Boromand,$^{\ast}$\textit{$^{a,b}$}   Alexandra Signoriello,\textit{$^{c}$}  Janna Lowensohn,\textit{$^{d}$} Carlos S. Orellana,\textit{$^{d}$} Eric R. Weeks,\textit{$^{d}$} Fangfu Ye,\textit{$^{b,e}$}  Mark D. Shattuck,\textit{$^{f}$} and Corey S. O'Hern\textit{$^{a,c,g,h}$}} \\

\includegraphics{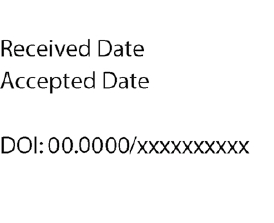} & \noindent\normalsize{We perform
computational studies of jammed particle packings in two dimensions
undergoing isotropic compression using the well-characterized soft
particle (SP) model and deformable particle (DP) model that we
developed for bubbles and emulsions. In the SP model,
circular particles are allowed to overlap, generating purely
repulsive forces.  In the DP model, particles minimize their
perimeter, while deforming at fixed area to avoid overlap during
compression.  We compare the structural and mechanical
properties of jammed packings generated using the SP and DP
models as a function of the packing fraction $\rho$, instead of
the reduced number density $\phi$.  We show that near jamming onset
the excess contact number $\Delta z=z-z_J$ and shear modulus ${\cal
G}$ scale as $\Delta \rho^{0.5}$ in the large system limit for
both models, where $\Delta \rho = \rho-\rho_J$ and
$z_J \approx 4$ and $\rho_J \approx 0.842$ are the values at jamming
onset. $\Delta z$ and ${\cal G}$ for the SP and DP models begin to
differ for $\rho \gtrsim 0.88$. In this regime, $\Delta z \sim {\cal
G}$ can be described by a sum of two power-laws in $\Delta \rho$,
i.e. $\Delta z \sim {\cal G} \sim C_0\Delta \rho^{0.5} +C_1\Delta
\rho^{1.0}$ to lowest order. We show that the ratio $C_1/C_0$ is
much larger for the DP model compared to to that for the SP model.
We also characterize the void space in jammed packings as a function
of $\rho$. We find that the DP model can 
describe the formation of Plateau borders as $\rho \rightarrow 1$. 
We further show that the results for $z$ and
the shape factor ${\cal A}$ versus $\rho$ for the DP model agree
with recent experimental studies of foams and emulsions.}
\\

\includegraphics{dates} & \\

\end{tabular}

 \end{@twocolumnfalse} \vspace{0.6cm}

  ]

\renewcommand*\rmdefault{bch}\normalfont\upshape
\rmfamily
\section*{}
\vspace{-1cm}


\footnotetext{\textit{$^{a}$~Department of Mechanical Engineering and Materials Science,
Yale University, New Haven, Connecticut, 06520, USA; E-mail: arman.boromand@yale.edu. }}
\footnotetext{\textit{$^{b}$~Beijing National Laboratory for Condensed Matter Physics and CAS Key Laboratory of Soft Matter Physics,
Institute of Physics, Chinese Academy of Sciences, Beijing, China.}}
\footnotetext{\textit{$^{c}$~Program in Computational Biology and Bioinformatics,
Yale University, New Haven, Connecticut, 06520, USA}}
\footnotetext{\textit{$^{d}$~Department of Physics, Emory University, Atlanta, Georgia 30322, USA.}}
\footnotetext{\textit{$^{e}$~School of Physical Sciences, University of Chinese Academy of Sciences, Beijing, China.}}
\footnotetext{\textit{$^{f}$~Benjamin Levich Institute and Physics Department,
The City College of New York, New York, New York 10031, USA.}}
\footnotetext{\textit{$^{g}$~Department of Physics, Yale University, New Haven, Connecticut, 06520, USA. }}
\footnotetext{\textit{$^{h}$~Department of Applied Physics, Yale University, New Haven, Connecticut, 06520, USA; E-mail: corey.ohern@yale.edu. }}

\rmfamily 


\section{Introduction}
\label{intro}

Soft materials, such as grafted core-shell particles, dendrimers, star
polymers, emulsions, foams, and hydrogels, are a class of materials
for which their microstructure can be altered by external fields,
applied deformation, and thermal fluctuations at room
temperature..~\cite{deg_revmod, weitz_book} The ability to vary
particle microstructure enables the design of soft materials with
novel functional properties and processing capabilities. Molecular
architecture, surface interactions, and deformability of soft
particles can be harnessed to develop novel soft composites with
optimized energy absorption, self-healing behavior, high mechanical
strength, and other desirable properties.\cite{lyklema,
  mattsson_nature, malescio_nature,velas_rev} In addition, many
biological systems such as biofilms~\cite{acemel_scirep}, cell
aggregates~\cite{bi_nature}, and tissues~\cite{mongera_nature} can be
considered as collections of soft and deformable particles. 

The interactions between soft particles, e.g. the softness, range, and
strength of the attraction and repulsion between soft particles is
controlled by their composition and microstructure. In turn, the
interactions between soft particles determine the collective
mechanical and rheological properties of packings of soft particles.
Significant challenges remain in understanding the influence of
particle microstructure and interactions on the macroscopic properties
of soft matter systems. In this article, we study the role of particle
deformability in determining the structural and mechanical properties
of packings of quasi-2D emulsions, modeled as collections of purely
repulsive, deformable particles at and above the jamming transition.

Systems composed of soft, frictionless particles, such as foams and
emulsions, can jam, or develop a non-zero static shear modulus ${\cal G}$,
when they are isotropically compressed to packing fractions that
approach random close packing $\phi_{J}$.~\cite{bonn_prl,dennin_pre,mason_prl}  For $\phi < \phi_J$,
packings of purely repulsive, frictionless spherical particles have an
insufficient number of interparticle contacts for them to be
mechanically stable. As a result, the packings exist at zero pressure
($p=0$) and are fluid-like, and particle rearrangements cost zero
energy.~\cite{liu_rev}  When compressed to $\phi_J$, the packings develop a connected
interparticle contact network with an isostatic number of contacts per
particle $z_J=N_c/N$, where $N_c = 2N'-1$, $N'=N-N_r$, $N$ is the total 
number of particles, and $N_r$ is
the number of rattler particles with less than $2$ contacts, for
frictionless circular particles in two spatial dimensions (2D) with
periodic boundary conditions.~\cite{witten_pre, corey_pre}  For $\phi>\phi_J$, the packings become
solid-like with $z>z_J$ and a nonzero shear modulus ${\cal G}>0$.~\cite{corey_prl, corey_pre,vanhecke_rev}

\begin{figure}[t]
\centering
  \includegraphics[trim={0.1cm 0.1cm 0.1cm 0.1cm},clip, width=0.48\textwidth]{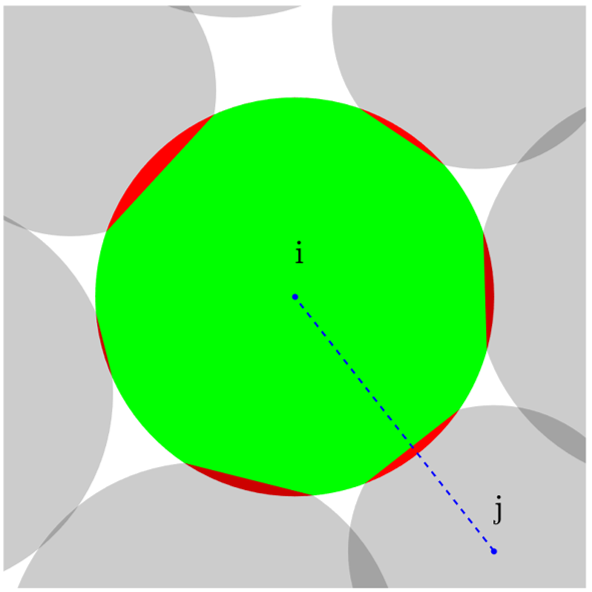}
\caption{A configuration of overlapping disks (e.g. dark grey regions 
between overlapping light grey particles) at $\phi = 1.0$ for the 
SP model. For the central green particle $i$ that overlaps several adjacent 
particles, we can define a more 
realistic packing fraction $\rho$ for the SP model, by associating half 
of each overlap to particle $i$ and the other half to each particle $j$ 
(red region) that overlaps $i$. The modified shape parameter for particle $i$ 
can then be obtained by calculating the perimeter and area of the 
green-colored shape. In this example, ${\cal A}_i=1.025$ and $\rho = 0.95$.}
\label{fig.0}
\end{figure}

A number of computational studies have been performed to investigate
the structural and mechanical properties of compressed foams and
emulsions in the solid-like regime $\phi > \phi_J$.~\cite{durian_prl, grest_prl, mobius_colloid, dunne_tf}  One of the most
frequently used models for characterizing their structural and mechanical 
properties is the soft particle (SP) model~\cite{durian_prl,corey_prl}, for which there is a potential energy cost proportional to the square of the overlap
between pairs of spherical particles, and no energy cost when the particles do
not overlap. These studies find that the excess contact number above
the isostatic number, 
\begin{equation} 
\label{contact_number} 
z-z_J \approx z_0^{\phi} (\phi - \phi_J)^{\alpha_0^{\phi}}, 
\end{equation} 
and the shear modulus,
\begin{equation} 
\label{shear_modulus} 
{\cal G} \approx {\cal G}_0^{\phi}(\phi -\phi_J)^{\beta_0^{\phi}},
\end{equation} 
obey power-law scaling relations with $\phi-\phi_J$, where the scaling
exponents $\alpha_0^{\phi}=\beta_0^{\phi}=0.5$ in the large-system
limit.~\cite{corey_prl, corey_pre, goodrich_prl} Further, these studies have found that the exponents do not
vary with the shape of the purely repulsive interaction potential and
are the same in 2D and 3D.~\cite{corey_pre, goodrich_prl, grest_prl, vanhecke_rev}.

In Eqs.~\ref{contact_number} and~\ref{shear_modulus}, we define the
packing fraction (or reduced number density) in 2D for packings of
circular disks as $\phi=\sum_{i=1}^N \pi \sigma_i^2/4A$, where
$\sigma_i$ is the diameter of disk $i$ and $A=L_x L_y$ is the area of
the simulation box with edge lengths $L_x=L_y$ in the $x$- and
$y$-directions. Note that when using this definition of $\phi$, the area of 
particle overlaps is multiply-counted. A
positive feature of the SP model is its simplicity, however, a
negative aspect is that the particles do not conserve area when the
packings are compressed above jamming onset.

There have also been a number of experimental~\cite{desmond_sm, katgert_epl, jasna_prl, bare_arxiv}
studies (as well as computational studies~\cite{mobius_colloid, dunne_tf}) of the
structural and mechanical properties of compressed foams and emulsions
in 2D and 3D. These studies also find power-law scaling of the excess
contact number,
\begin{equation}
\label{contact_number_rho}
z-z_J \sim (\rho-\rho_J)^{\alpha'},
\end{equation}
where $\alpha'$ occurs in the range $0.4$ to $1$ depending on the
particular study. The power-law scaling of $z-z_J$ is measured versus
$\rho-\rho_J$, not $\phi-\phi_J$, where $\rho$ is the true packing
fraction of the system. For systems at jamming onset, $\rho_J =
\phi_J$, but $\phi > \rho$ when particles overlap in the SP
model. Also, $\phi \geq 1.0$ is allowed, whereas $\rho \leq 1.0$ is a
hard constraint. In addition, experimental studies of compressed
emulsions in 3D have shown that the shear modulus obeys power-law
scaling in $\rho-\rho_J$,
\begin{equation}
\label{shear_modulus_rho}
{\cal G} \sim (\rho - \rho_J)^{\beta'},
\end{equation}
but the scaling exponent $\beta' > \beta^\phi_0$.~\cite{mason_prl, mason_pre,princen_jcis}     

In light of the discrepancies between the power-law scaling exponents
found in the experimental studies of compressed foams and emulsions
and those obtained from the computational studies of the SP model, we
employ the recently developed deformable particle (DP) model~\cite{arman_prl}for foams and
emulsions to understand how particle deformability affects the
packing fraction dependence of the structural and mechanical properties of 
jammed particle packings.

Other computational methods have been employed to model particle
deformability in soft matter systems, such as foams and
emulsions. There are two main classes of methods for modeling particle
deformability: Lattice Boltzmann~\cite{ sun_rheact, benzi_epl} and
particle-based methods~\cite{mobius_colloid, kahara_pre, kern_pre,
  honda, rognon_epe}. The lattice-based methods have typically focused
on two- or multi-phase modeling, whereas the DP model focuses only on
the shape degrees of freedom of the particles (i.e. bubbles or
droplets). Our work on the DP model differs from the previous studies
using particle-based methods. First, the work by Rognon, {\it et
al}.~\cite{rognon_epe} has been limited to small systems composed of
$2-5$ particles. Second, the froth model by Kern, {\it et
al}.~\cite{kern_pre} is limited to the dry regime, where the true
packing fraction approaches unity. In contrast, the DP model can be
used to study a wide range of packing fractions, from values where the
particles are out of contact to confluent systems. The model proposed
by Kahara {\it et al}.~\cite{kahara_pre} is most similar to the DP
model. However, by modeling the pressure of the carrier fluid, their
study is limited to the wet regime. In addition, their studies have
focused on the rheological properties of bubbles during shear. In
contrast, this article will focus on the structural and mechanical
properties of jammed deformable particles generated during isotropic
compression.

A central assumption of the SP model is that the particles remain
spherical as particle overlap increases when the system is compressed
above jamming onset.  For this reason, studies that employ the SP
model typically quantify the system properties as a function of $\phi$ instead
of the true packing fraction $\rho$~\cite{desmond_sm, katgert_epl}.
\begin{figure*}[t]
\centering
  \includegraphics[trim={0.1cm 0.1cm 0.1cm 0.1cm},clip, width=1.0\textwidth]{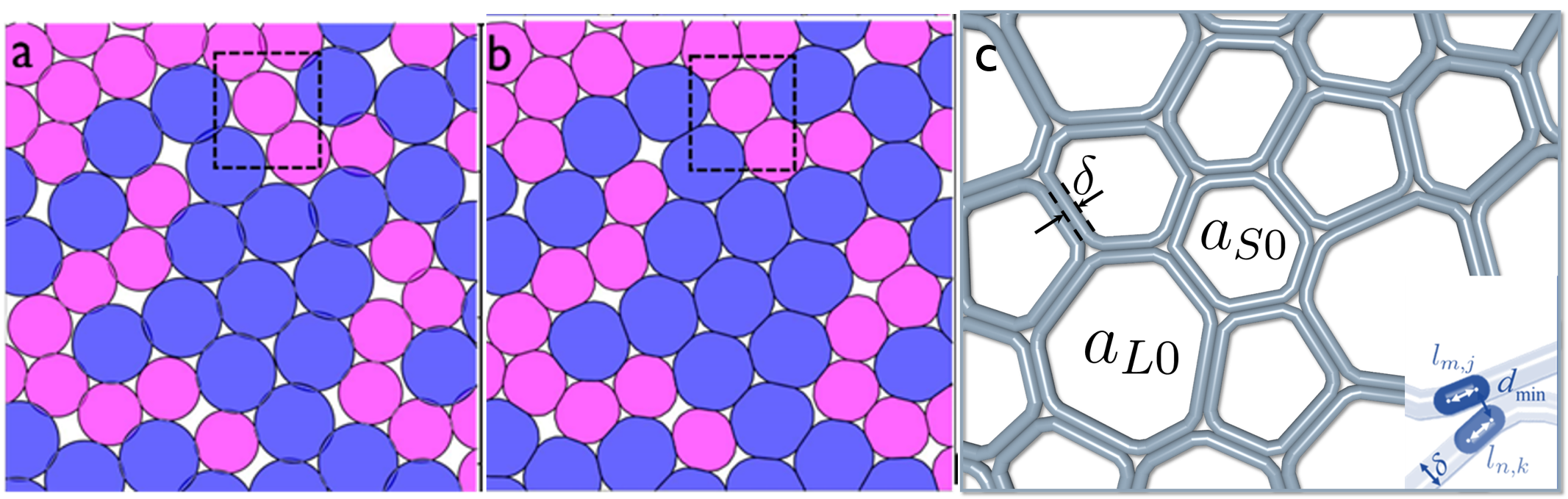}
  \caption{Similar jammed packings of $N=64$ bidisperse disks [half large (blue) 
and half small (pink) disks with diameter ratio $\sigma_L/\sigma_S=1.4$] 
with packing fraction $\rho \approx 0.92$ generated using the (a) soft particle
(SP) model and (b) deformable particle (DP) model for foams and emulsions. 
The average shape parameter is $\langle {\cal A} \rangle \approx 1.01$ 
and $1.03$ for the SP and DP models, respectively. The dashed boxes 
highlight extra contacts that form in the packing of deformable particles 
compared to the soft particle packing during compression. (c) Close-up of a jammed configuration of $N=128$ deformable
particles at $\rho = 0.99$ and shape parameter $\langle {\cal A} \rangle = 
1.06$. Each 
deformable particle is a
collection of $N_v$ interconnected circulo-lines with width
$\delta$. The system includes $N/2$ large particles with $N_v = 17$ 
and $N/2$ small particles with $N_v = 12$. The preferred areas of the 
particles (in 
Eq.~\ref{tote2d}) are $a_{L0}$ and $a_{S0}$, for the large and small 
particles, respectively, with $a_{L0}/a_{S0} \approx 2.0$. 
The inset shows two interacting deformable particles $m$ and $n$. $U_{\rm 
int}$ is proportional to $(\delta - d_{\rm min})^2$, where $d_{\rm min}$ is 
the minimum distance between overlapping circulo-lines $j$ and $k$ on 
deformable particles $m$ and $n$.}
\label{fig.1}
\end{figure*}
 Even though the SP model does not
conserve particle area as the system is compressed, one can also measure
the structural and mechanical properties as a function of the true
packing fraction $\rho$ when using the SP model by attributing half of an
overlap between particles $i$ and $j$ to particle $i$ and the other half to
particle $j$. (See Fig.~\ref{fig.0}.)

A key feature in jammed packings of foams and emulsions is that the
particles maintain their area (volume) during compression over the
full range of packing fraction.  Bubbles in foams and emulsion
droplets can deform, become non-spherical, and form additional
contacts that do not occur in the SP model at a comparable value of
packing fraction as that shown in Fig.~\ref{fig.1} (a) and (b). In
this article, we show that the soft particle and deformable particle
models show similar results for the scaling of the excess contact
number and shear modulus versus $\rho-\rho_J$ for packing fractions
close to jamming onset.  However, for larger $\rho$, we find that
$z-z_J$ and ${\cal G}$ for the SP and DP models begin to differ
significantly. In this regime, $z(\rho)$ for the deformable particle
model is similar to that found for experimental studies of compressed
emulsions and foams in 2D. We also study the geometric properties of
the void space of jammed packings as a function of $\rho$. We show
that unlike the SP model, the DP model is able to recapitulate the
formation of Plateau borders~\cite{cohen_pre, weair_book, weair_sm,
  gay_epe}, where bubble edges have a relative orientation of
$120^{\circ}$ and form a void with shape factor ${\cal A} \approx
4.87$, near confluence. (The shape factor ${\cal A} = p^2/4\pi a$,
where $p$ is the perimeter and $a$ is the area of the void
space~\cite{arman_prl, bare_arxiv}.)

The remainder of the article is organized as follows. In
Sec.~\ref{method}, we describe the soft particle and deformable
particle models for 2D compresed foams and emulations, and the
isotropic compression protocol that we employ to numerically generate
jammed packings. In Sec.~\ref{results}, we compare the results for the
structural and mechanical properties of jammed packings using the SP
and DP models. We show the variation of $\rho$ with $\phi$ above
jamming onset and quantify the decrease in the average area of the
particles as a function of increasing packing fraction for the SP
model.  We measure $\rho$ versus the shape factor ${\cal A}$ to
determine at what shape factor the SP and DP models reach
confluence. We then show the power-law scaling results for $z-z_J$ and
${\cal G}$ versus $\rho - \rho_J$ for the SP and DP models. We also
characterize the connected void regions, by measuring the number,
size, and shape of the voids as a function of packing fraction for the
SP and DP models.  In Sec.~\ref{conclusions}, we summarize the
results for the SP and DP models, compare the results for the DP model to those from recent
experiments on compressed foams and emulsions, and discuss
future research directions.


\section{Simulation Methods}
\label{method}

In this article, we study the structural and mechanical properties of
isotropically compressed jammed packings of $N$ purely repulsive,
frictionless bidisperse particles in 2D using the soft particle and
deformable particle models. For both models, the simulation cell is
square with periodic boundaries in both directions.

\subsection {Soft Particle Model}
\label{ssc-sp}

For the soft particle model, pairs of circular disks $i$ and $j$ interact 
via the purely repulsive pairwise potential: 
\begin{eqnarray}
\label{tote2s}
U_{SP}(r_{ij}) & = & \frac{\epsilon}{2} \left(1-\frac{r_{ij}}{\sigma_{ij}}\right)^2 \Theta \left(1-\frac{r_{ij}}{\sigma_{ij}}\right),
\end{eqnarray}
where $\epsilon$ is the characteristic energy scale of the
interaction, $\sigma_{ij}$=$(\sigma_i + \sigma_j)/2$ is the average
diameter and $r_{ij}$ is the center-to-center separation between disks
$i$ and $j$, and $\Theta(.)$ is the Heaviside step function that sets the
interaction potential to zero when disks $i$ and $j$ do not
overlap. We focused on systems composed of $N/2$ large and $N/2$ small
disks with equal mass $m$ and diameter ratio, $\sigma_L/\sigma_S= 1.4$
to avoid crystallization. The total potential energy for the SP model is  
$U_{SP}=\sum_{i>j} U_{SP}(r_{ij})$ and the stress tensor 
is given by 
\begin{equation}
\label{stress}
\Sigma_{\mu \nu} = A^{-1} \sum^N_{i>j} f_{ij\mu} 
r_{ij\nu}, 
\end{equation}
where $\mu$, $\nu = x$ and $y$ and ${\vec f}_{ij} = -{\vec
  \nabla}_{r_{ij}} U_{SP}(r_{ij})$. To measure the shear modulus
${\cal G}$, we apply an infinitesimal affine shear strain $\gamma$ to the
$x$-positions of the particle centers, $x'_i = x_i + \gamma y_i$, and
measure the resulting shear stress $\Sigma_{xy}$. We then calculate
the shear modulus ${\cal G} = -d\Sigma_{xy}/d\gamma$ (at fixed area).
For the SP model, we measure energy, length, and stress in units of
$\epsilon$, $\sigma_{LS}$, and $\epsilon/\sigma^2_{LS}$.

\subsection {Deformable Particle Model}
\label{ssc-edp}

To model bubbles and droplets, we consider the deformable particle
model~\cite{arman_prl} for foams and emulsions with a potential
energy that includes the following three terms:
\begin{eqnarray}
\label{tote2d}
U_{DP} & = &\gamma \sum_{m=1}^N \sum_{i=1}^{N_v} l_{m,i} +
\frac{k_a}{2} \sum_{m=1}^{N} (a_m-a_{m0})^2\\
& + & U_{\rm int}. \nonumber
\nonumber
\end{eqnarray}
Each deformable ``particle'' (indexed by $m=1,\ldots,N$) is modeled as
a polygon with $N_v$ circulo-line edges to represent $N_v-1$ shape
degrees of freedom.  The circulo-lines have width
$\delta$~\cite{kyle_pre} and are indexed by $i=1 \ldots,N_v$. (See
Fig.~\ref{fig.1}c).
\begin{figure*}[t]
\centering
 \includegraphics[trim={0.2cm 0.2cm 1.5cm 0.5cm},clip, width=1.0\textwidth]{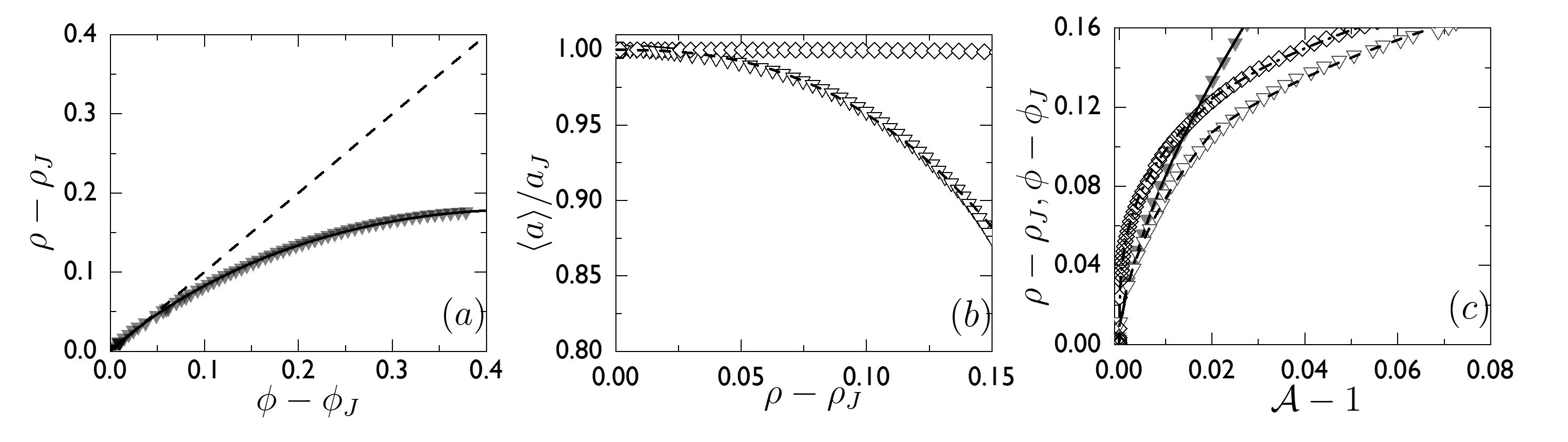}
\caption{(a) The packing fraction $\rho-\rho_J$ versus the reduced number density
$\phi-\phi_J$ for static packings of $N=32$ bidisperse disks using the SP model. 
The dashed line is $\rho=\phi$. The solid line indicates 
$\rho-\rho_J = C (\phi-\phi_J)^{0.5}[1-(\phi-\phi_J)]$, where $C={\cal O}(1)$. (b) Average particle area $\langle a \rangle$ normalized by the value at jamming 
onset $a_J$ as a function of $\rho - \rho_J$ for the SP (downward triangles) 
and DP (diamonds)
models with $N=32$. The dashed line has the form $a/a_J -1 \propto (\rho-\rho_J)^\zeta$, with $\zeta \approx 2.5$ to capture the large $\rho-\rho_J$ 
behavior. (c) The packing fraction $\rho-\rho_J$ and reduced number density
$\phi-\phi_J$ versus the shape parameter, ${\cal A}-1$, for the DP (diamonds) and SP (triangles) models. $\phi=1$ occurs at ${\cal A} = 1.03$ for the SP model 
(filled triangles). In contrast, $\rho=1$ at ${\cal A}> 1.10$ for the SP 
model (open triangles). Packings generated using the DP model reach confluence at  
${\cal A} \approx 1.07$. The dashed-dotted line has the form 
$\rho - \rho_J \propto ({\cal A} - 1)^\omega$ with $\omega \approx 0.3$ 
for the DP model, which captures the large $\rho-\rho_J$ 
behavior. The dashed line through the SP model data has a similar form, but with two scaling regimes: 
one at small ${\cal A}-1$ with $\omega \approx 0.5$ and one at large 
${\cal A}-1$ with $\omega \approx 0.3$. The solid line has the form $\phi-\phi_J \propto ({\cal A}-1)^{\lambda}$
with $\lambda \approx 0.67$ for the SP model.}
\label{fig.3}
\end{figure*}
We consider $N/2$ large particles with $N_v = 17$ and $N/2$ small
particles with $N_v = 12$, and $a_{L0}/a_{S0} = (17/12)^2 \approx
2.0$, which is similar to the area ratio of the large and small disks
in the SP model.  We have also studied DP packings with larger numbers
of vertices (while maintaining $a_{L0}/a_{S0} \approx 2.0$), and the
structural and mechanical properties are similar to those for $N_v =
17$ and $12$ for the large and small particles, respectively.  The
location of the $i$th circulo-line in particle $m$ is ${\vec
  v}_{m,i}$, and the bond vector ${\vec l}_{m,i} = {\vec v}_{m,i+1} -
{\vec v}_{m,i} = l_{m,i} {\hat l}_{m,i}$ connects circulo-lines $i+1$
and $i$.

The first term in $U_{DP}$ is proportional to the total length of the
interface, i.e. the perimeter, $p_m=\sum_{i=1}^{N_v}{l_{m,i}}$ of the
$m$th particle with a proportionality constant equal to the line
tension $\gamma$. The second term is quadratic in $a_m$ with a minumum
at $a_{L,S 0}$, which penalizes deviations in area from the reference 
value $a_{L,S 0}$. Here, we study $k_a a_{L,S0}^2 > 10^3$, which implies that  
the fluctuations in the particle areas are $\lesssim 10^{-3}$. We
characterize the shape of the deformable particles by calculating the
particle shape parameter ${\cal A}_m = p_m^2/4 \pi a_m$, which equals
$1$ for circular disks and is greater than $1$ for all non-spherical
shapes~\cite{bi_prx,arman_prl}.

Note that for the DP model for foams and emulsions, we remove the
constraint on the elastic interface~\cite{arman_prl}, i.e. the
preferred bond length, $l_{m,i} = 0$. As a result, the spacing between
the vertices on a given deformable particle can change as they
interact with vertices on neighboring particles.  This allows us to
correctly model the formation of elongated edges when deformable
particles make contacts (such that $l_{m,i}> 0$), as well as model the
formation of Plateau borders~\cite{weair_book,cohen_pre} with $l_{m,i}
\rightarrow 0$.

The third term, $U_{\rm int}$, penalizes overlaps between deformable 
particles by including purely repulsive interactions between pairs 
of contacting circulo-lines on neighboring deformable particles: 
\begin{eqnarray}
\label{int}
U_{\rm int} & = & \sum_{m=1}^N \sum_{n>m}^N\sum_{j=1}^{N_v}\sum_{k=1}^{N_v}\frac{\epsilon_r}{2}\left(1-\frac{d_{\rm min}}{\delta}\right)^2\\
& \times &\Theta\left(1-\frac{d_{\rm min}}{\delta}\right),
\nonumber
\end{eqnarray}
where $\epsilon_r$ gives the strength of the repulsive interactions, $d_{\rm
  min}$ is the minimum distance between circulo-lines $j$ and
$k$ on contacting deformable particles $m$ and $n$, and
$\Theta(.)$ ensures that there is no interaction when the
circulo-lines on different particles are out of contact. The stress tensor for packings of deformable particles is 
obtained using $\Sigma_{\mu \nu} = A^{-1} \sum_{i=1}^N f_{i \mu}^{\rm ext} 
r^c_{i \nu}$, where ${\vec f}^{\rm ext}_{i} = -{\vec \nabla}_{r_{i}}
U_{\rm int}$ is the force on particle $i$ arising from $U_{\rm int}$ 
and ${\vec r}_{i}$ is the position of the centroid of particle $i$. 

\begin{figure}
\centering
\includegraphics[trim={0.1cm 0.2cm 0.0cm 0.2cm},clip, width=0.5\textwidth]{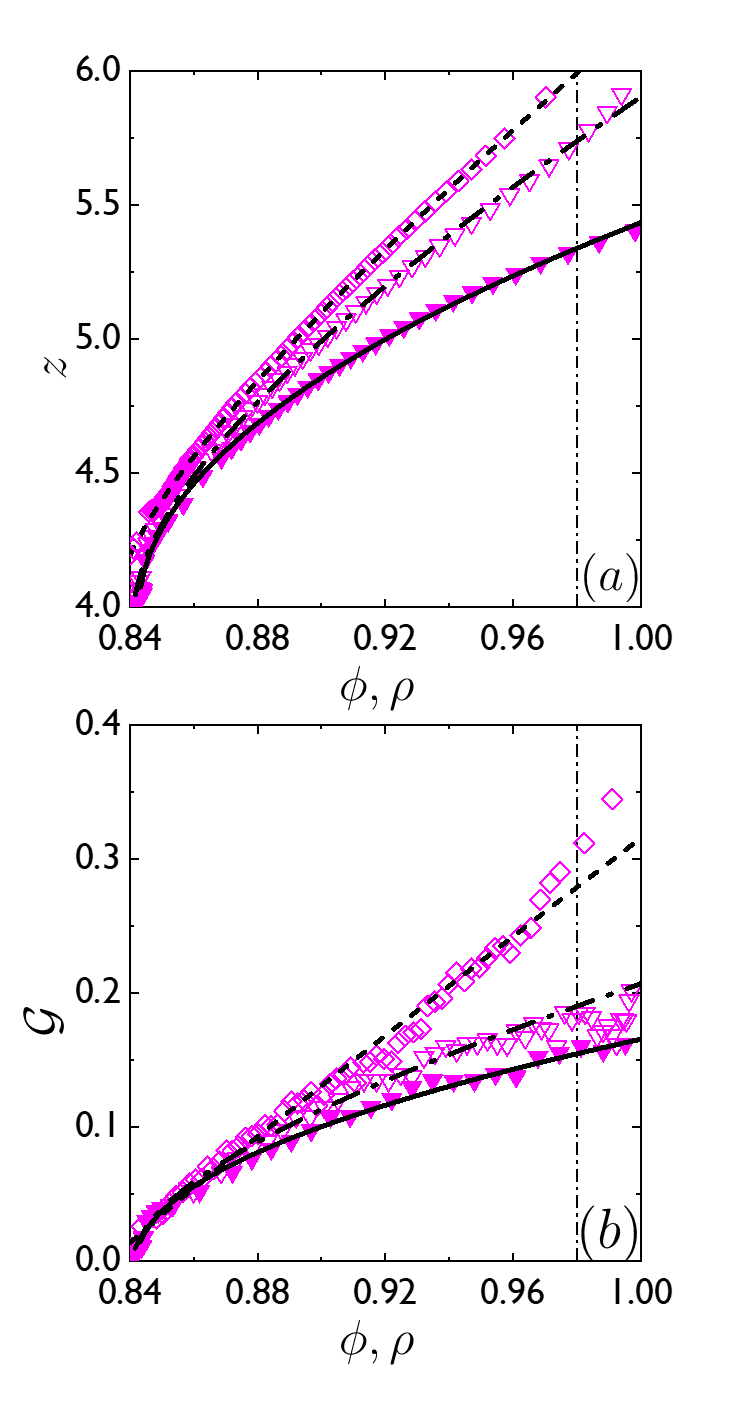}
\caption{(a) The number of contacts per particle $z$ and (b) shear 
modulus ${\cal G}$ versus the true 
packing fraction $\rho$ for jammed 
packings of  $N=256$ particles generated using the SP 
(open downward triangles) and
DP models (open diamonds). We also 
show $z$ and ${\cal G}$ versus the reduced 
number density $\phi$ for the SP model (filled downward triangles). The dashed and dashed-dotted lines 
are fits of $z$ and ${\cal G}$ versus $\Delta \rho$ for the DP and SP models, respectively, using the forms in Tables~\ref{table:nonlin1} and \ref{table:nonlin2}.
The solid lines are fits of $z(\Delta \phi)$ and ${\cal G}(\Delta \phi)$ to the 
forms in Tables~\ref{table:nonlin1} and \ref{table:nonlin2} for the SP model. The dashed-dotted vertical lines 
indicate the packing fraction above which the measurements start to 
deviate from the power-law scaling forms in Eqs.~\ref{zrho} and ~\ref{Grho}.}
\label{fig.4}
\end{figure}
To measure the shear modulus ${\cal G}$, we apply an infinitesimal
affine shear strain $\gamma$ to the $x$-positions of the
$i=1,\ldots,N_v$ circulo-lines on each particle $m$, $v_{xm,i}' =
v_{xm,i} + \gamma v_{ym,i}$, and measure the resulting shear stress
$\Sigma_{xy}$. We then calculate the shear modulus ${\cal G} =
-d\Sigma_{xy}/d\gamma$ (at fixed area).  For the DP model, we measure
energy, length, and stress in units of $\epsilon_r$, $l$, and
$\epsilon_r/l^2$, where $l=(\sqrt{a_{S0}}+\sqrt{a_{L0}})/2$. The
structural and mechanical properties of DP packings at jamming onset
do not depend on the parameters $\gamma$, $k_a$, and $\epsilon_r$.
However, above jamming onset, the properties can depend on these
parameters. We focused on the parameter regime, $\epsilon_r >
k_a(a_{L,S0})^2 > \gamma \langle p_m\rangle$, which is typical for
foams and emulsions.

\subsection {Isotropic Compression Packing Protocol}
\label{protocol}

The protocol to generate jammed packings is similar for the SP and DP
models. The protocol proceeds in two stages. For each initial
condition, we first identify the packing fraction $\rho_J = \phi_J$ at jamming onset.  For the SP model, the system is
initialized using random locations for the disks at $\rho=0.20$. For
the DP model, we place the particle centers randomly at $\rho=0.20$
and then position the $N_v$ circulo-lines equally spaced around each
particle center. We successively isotropically compress the system (by
decreasing the size of the simulation cell) using small packing
fraction increments ($\delta \phi$=$10^{-4}$ for the SP model and
$\delta \rho = 10^{-4}$ for the DP model) and minimize the total
potential energy per particle, $U_{SP}/N$, for the SP model (or $U_{DP}/(N N_v)$
for the DP model) after each compression step using over-damped
molecular dynamics simulations until the kinetic energy per particle
satisfies $K/N < 10^{-20}$ for the SP model and the total kinetic
energy per circulo-line $K/(N N_v) < 10^{-20}$ for the DP model. If
$U_{SP}/N$ or $U_{DP}/(N N_v)$ is ``zero'' (i.e. $U_{SP}/N <10^{-15}$ or 
$U_{DP}/(N N_v) <
10^{-15}$) after minimization, the system is subsequently compressed. If $U_{SP}/N$ or
$U_{DP}/(N N_v)$ is nonzero, i.e. there are finite particle overlaps and
$U_{SP}/N > 10^{-13}$ or $U_{DP}/(N N_v) > 10^{-13}$, after minimization, the
system is subsequently decompressed. The increment by which the packing fraction is
changed at each compression or decompression step is gradually
decreased. 

We terminate the process of finding the onset of jamming at
$\rho_J$ or $\phi_J$ when the system satisfies $10^{-13} < U_{SP}/N < 10^{-16}$ and
$K/N < 10^{-20}$ for the SP model or $10^{-13} < U_{DP}/(N N_v) < 10^{-16}$
and $K/(N N_v) < 10^{-20}$ for the DP model.  This process yields
mechanically stable packings at jamming onset.

The second stage of the protocol involves sampling the system at set
of packing fractions $\rho - \rho_J >0$ above jamming onset with adjacent
values separated by $\delta \rho = 10^{-4}$ for the DP model or
a set of $\phi - \phi_J >0$ with adjacent values separated by
$\delta \phi = 10^{-4}$ for the SP model. Ensemble averages are
obtained by averaging over systems at fixed $\phi - \phi_J$ or $\rho -
\rho_J$ for the SP and DP models, respectively, where $\phi_J$ and $\rho_J$
are determined separately for each initial condition. The 
distribution of $\phi_J$ and $\rho_J$ for the SP and DP models are shown 
for several system sizes in the Appendix.

\section{Results}
\label{results}

Below, we compare the results for the structural and mechanical
properties of particle packings as a function of packing fraction
above jamming onset obtained using the soft particle and deformable
particle models. 
\begin{figure*}[t]
\centering
\includegraphics[trim={0.1cm 0.2cm 0.0cm 0.0cm},clip, width=\textwidth]{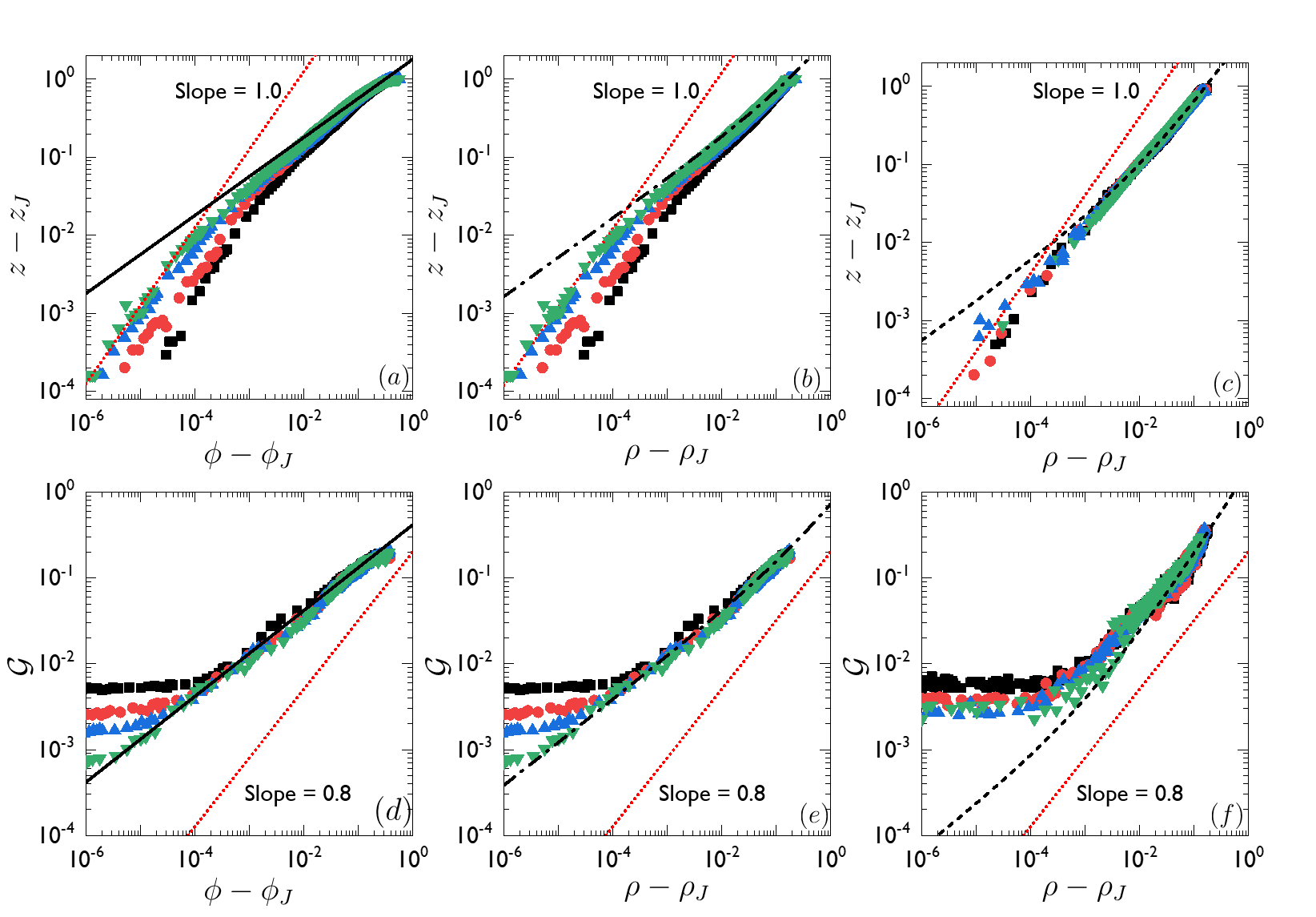}
\caption{Panels (a) and (d) show the excess contact number $z-z_{J}$  and 
shear modulus ${\cal G}$ plotted versus $\Delta \phi$ for the SP model. 
(b) and (e) show the same data in panels (a) and (d), except plotted versus 
$\Delta \rho$. Panels (c) and (f) show $z-z_J$ and ${\cal G}$ versus 
$\Delta \rho$ for the DP model. Each symbol represents different system sizes:
$N = 32$ (squares), $64$ (circles), $128$
(upward triangles), and $256$ (downward triangles). The dotted lines in 
panels (a)-(c) have slope equal to $1$. The solid line in (a), 
dashed-dotted line in (b), and dashed line in (c) are the same fits  
to the data as in Fig.~\ref{fig.4} (a). 
The dotted lines in panels (d)-(f) have slope equal to $0.8$.  The solid 
line in (d), dashed-dotted line in (e), and dashed line in (f) are the
same fits to the data as in Fig.~\ref{fig.4} (b).}
\label{fig.5}
\end{figure*}
These studies allow us to investigate the effect of
particle deformability on the structural and mechanical properties of
jammed solids. As discussed in Sec.~\ref{intro}, there are several key
differences between the SP and DP models. For example, the SP model
allows overlap between particles and concomitant decreases in the
particle area as the system is compressed above jamming onset.
In contrast, the particles in the DP model {\it deform} to prevent
interparticle overalps, and thus they maintain their areas, and do not
remain circular in shape.

A method to minimize the effects of the loss of particle area for the
SP model is to quantify the structural and mechanical properties of
jammed packings generated using the SP model as a function of the true
packing fraction $\rho$, not $\phi$.  To measure $\rho$ at each $\phi
> \phi_J$, we need to subtract from $\phi A$ the muliply-counted areas of
overlapping disks. For $\phi<1.2$, we only need to consider overlaps
between pairs of disks, i.e. subtract off the area of each lens
between pairs of overlapping disks.  (See Fig.~\ref{fig.0}.) In this
case, the true packing fraction is
\begin{equation}
\label{true_rho}
\rho = \sum_{i=1}^N \frac{\pi \sigma_i^2}{4A} - \frac{1}{A} \sum_{i>j}^N a_{ij}^{\rm ov},
\end{equation}
where 
\begin{equation}
\label{lens_area}
a_{ij}^{\rm ov}=\frac{\sqrt{(-r_{ij}+\sigma_{ij})(r_{ij}-{\overline \sigma}_{ij})(r_{ij}+{\overline \sigma}_{ij})(r_{ij}+\sigma_{ij})}}{2}
\end{equation}
is the area of the lens between overlapping disks $i$ and $j$ and
${\overline \sigma}_{ij} = (\sigma_i-\sigma_j)/2$. For $\phi > 1.2$,
the lens between overlapping disks $i$ and $j$ can overlap with the
lens of other overlapping pairs of disks, which modifies
Eq.~\ref{true_rho}.

In Fig.~\ref{fig.3} (a), we plot the deviation in the true packing
fraction from that at jamming onset, $\Delta \rho = \rho - \rho_J$,
versus the deviation in the reduced number density $\phi$ from the
value at jamming onset, $\Delta \phi = \phi-\phi_J$, for jammed packings
generated using the SP model.  On a linear scale, $\rho \approx \phi$
for $\phi \lesssim 0.88$. More generally, we find
\begin{equation}
\label{exact}
\Delta \rho \approx C(\Delta \phi - \Delta \phi^{1.5}), 
\end{equation}
where $C$ is weakly dependent on $\phi$. (See Figs.~\ref{app.2} (a)
and (b) in the Appendix.)  Note that Eq.~\ref{exact} is exact when
higher-order overlaps (i.e. the lens from overlapping disks $i$ and
$j$ overlaps with the lens from other overlapping disks) do not
occur. To study the structural properties of the SP model near
confluence, $\rho \rightarrow 1$, we also considered cases where three
disks mutually overlap. The true packing fraction becomes
\begin{equation}
\label{3body_rho}
\rho = \sum_{i=1}^N \frac{\pi \sigma_i^2}{4A} - \frac{1}{A} \sum_{i>j}^N a_{ij}^{\rm ov}+ \frac{2}{A} \sum_{k>i,j}^N a_{ijk}^{\rm ov},
\end{equation}
where $a_{ijk}^{\rm ov}$ is the area of the Reuleaux triangles that
form when three disks mutually overlap.  Using this approximation, we
find that $\rho \rightarrow 1$ near $\phi \approx 1.24$.

As discussed in the introduction, for packings generated using the SP
model, the area of the particles decreases with increasing packing
fraction above jamming onset. We calculate the average area of the
particles (normalized by the average at jamming onset) $\langle
a\rangle/a_J$ versus $\Delta \rho$ for packings generated using the SP
and DP models in Fig.~\ref{fig.3} (b).  On a linear scale, $\langle a
\rangle$ begins deviating significantly from $a_J$ for $\Delta \rho
\gtrsim 0.04$ for the SP model.  In the Appendix, we show that $a/a_J
-1 \propto \Delta \rho^\zeta$ with an exponent $\zeta \approx 1.5$ at
small $\Delta \rho$ and $\zeta \approx 2.5$ at large $\Delta \rho$ for
the SP model. In contrast, $\langle a \rangle \approx a_J$ over the
full range of $\Delta \rho$ for the DP model.

In Fig.~\ref{fig.3} (c), we quantify how the particles deform during
isotropic compression above jamming onset. In general, the packing
fraction increases with the shape parameter ${\cal A}-1$. In the
Appendix, we show that $\rho-\rho_J$ grows as a power-law in the
deviation of the shape parameter from that at jamming onset, $({\cal
  A}-1)^{\omega}$. At small ${\cal A}-1$, $\omega \approx 0.66$ and
$\approx 1.0$ for the SP and DP models, respectively. At large ${\cal
  A}-1$, $\omega \approx 0.3$ for both the SP and DP models.
$\phi-\phi_J$ for the SP model also grows as a power-law with ${\cal
  A}-1$, but much faster than $\rho-\rho_J$.

At what shape parameter do 2D foams and emulsions reach
confluence? Fig.~\ref{fig.3} (c) shows that the DP and SP models reach
confluence ($\Delta \rho \approx 0.16$) at different values of the
shape parameter, ${\cal A} \approx 1.07$ for the DP model and ${\cal
  A} >1.10$ for the SP model. Thus, we find similarites and
differences between the shapes of the particles for packings generated
using the SP and DP models as they are compressed above jamming
onset. An interesting similarity is that the packing fraction for both
the SP and DP models scales as $\rho-\rho_J \sim ({\cal
  A}-1)^{\omega}$, where $\omega \approx 0.3$ at large $\rho$.

Next, we compare the contact number $z$ and shear modulus ${\cal G}$
versus the packing fraction for the SP and DP models. On a linear
scale, which emphasizes the values at large packing fraction, we find
weak system-size dependence for $z$ and ${\cal G}$ for packings
generated via the SP and DP models. Note that for the deformable particle 
model, multiple circulo-lines on one deformable
particle can be in contact with multiple circulo-lines on another
deformable particle. These multiple circulo-line contacts are treated
as a single contact between two deformable particles. 

In Fig.~\ref{fig.4}, for both $z$ and ${\cal G}$, measured in packings
of $N=256$ particles, the results for the SP and DP models are similar
near jamming onset $\phi_J \approx \rho_J$. For $z$ and ${\cal G}$,
the results for the SP and DP models begin to deviate near $\rho
\approx 0.88$. We find that more interparticle contacts form as the
packings are compressed above jamming onset for the DP model, compared
to that for the SP model.  As a result, the shear modulus grows more
rapidly with $\rho$ for packings generated using the DP model.
We also show the best fits of $z$ and ${\cal G}$
for the SP model 
to the power-law scaling form with
$\Delta \phi$ in Eqs.~\ref{contact_number} and~\ref{shear_modulus}. As
found previously, the scaling exponents
$\alpha_0^{\phi}=\beta_0^{\phi}\approx 0.5$ for the SP model.\cite{corey_pre, liu_rev}  By
plugging Eq.~\ref{exact} into Eqs.~\ref{contact_number}
and~\ref{shear_modulus}, we can convert $z(\Delta \phi)$ and ${\cal G}(\Delta
\phi)$ to $z(\Delta \rho)$ and ${\cal G}(\Delta \rho)$. To lowest order in
$\Delta \rho$, we find
\begin{equation}
\label{zrho}
z-z_J\approx z_0^{\rho}(\rho-\rho_J)^{\alpha_0^{\rho}} + z_1^{\rho}(\rho-\rho_J)^{\alpha_1^{\rho}} 
\end{equation}
and 
\begin{equation}
\label{Grho}
{\cal G}\approx {\cal G}_0^{\rho}(\rho-\rho_J)^{\beta_0^{\rho}} + {\cal G}_1^{\rho}(\rho-\rho_J)^{\beta_1^{\rho}} 
\end{equation}
for the SP model, where $\alpha_0^{\rho}\approx \beta_0^{\rho} \approx
0.5$ and $\alpha_1^{\rho} \approx \beta_1^{\rho} \approx 1.0$.  We
show fits of $z(\Delta \rho)$ and ${\cal G}(\Delta \rho)$ for the SP
model to Eqs.~\ref{zrho} and~\ref{Grho} as dashed-dotted lines in
Fig.~\ref{fig.4}.  The combination of the two power-laws in $\Delta
\rho$ with exponents $0.5$ and $1.0$ accurately describes the data for
$\Delta z$ and ${\cal G}$.  However, $\Delta z$ begins deviating from
Eq.~\ref{zrho} for $\rho \gtrsim 0.98$ near confluence. Moreover, we
find that Eqs.~\ref{zrho} and ~\ref{Grho} can be used to fit the data
for the DP model as well. The parameters for the fitting functions are
shown in Tables~\ref{table:nonlin1} and ~\ref{table:nonlin2}.

In Fig.~\ref{fig.5} (a) and (d), we present the excess contact number
$z-z_J$ and shear modulus ${\cal G}$ versus $\Delta \phi$ on
logarithmic axes for jammed packings using the SP model for system
sizes ranging from $N = 32$ to $256$. By plotting the data on
logarithmic scales, we can identify several different regimes in
$\Delta \phi$: 1) $\Delta \phi <10^{-3}$, 2) $10^{-3}< \Delta \phi <
0.2$, and 3) $\Delta \phi >0.2$. Regime $1$, where $\phi \approx
\phi_J$, is difficult to see on the linear scales shown in
Fig.~\ref{fig.4} (a).  In this regime, there is strong system-size
dependence and $\Delta z \sim \Delta \phi$ and ${\cal G} \sim \Delta
\phi^0$.~\cite{goodrich_prl,goodrich_pre}  As found previously, at intermediate $\Delta \phi$, $\Delta
z \sim \Delta \phi^{\alpha_1^{\phi}}$ and ${\cal G} \sim \Delta
\phi^{\beta_1^{\phi}}$, where $\alpha_1^{\phi} \approx \beta_1^{\phi} \approx 0.5$. The characteristic $\Delta \phi^*$ at
which the $\alpha$ and $\beta$ exponents cross-over from $1$ to $0.5$
and $0$ and $0.5$, respectively, scales as $\Delta \phi^* \sim
N^{-2}$.  (See Tables~\ref{table:nonlin1} and~\ref{table:nonlin2}.)
Thus, regime $2$ extends to smaller $\Delta \phi$ as the system size
increases. In regime $3$, at large $\Delta \phi$, we find that the
power-law scaling behavior of $\Delta z$ and ${\cal G}$ with $\Delta
\phi$ breaks down. 

Using Eq.~\ref{exact}, we can convert the data from functions of
$\Delta \phi$ to functions of $\Delta \rho$ for the SP model. We show
$\Delta z(\Delta \rho)$ and ${\cal G}(\Delta \rho)$ in
Fig.~\ref{fig.5} (b) and (d).  We find two different regimes in
$\Delta \rho$. At low $\Delta \rho$, $z-z_J \sim \Delta \rho$ and
${\cal G} \sim \Delta \rho^0$. At intermediate and large $\Delta
\rho$, the forms for $z-z_J$ and ${\cal G}$ include a sum of two
power-laws as shown in Eqs.~\ref{zrho} and~\ref{Grho}. The
characteristic packing fraction that separates the small and large
$\Delta \rho$ regimes also scales as $\Delta \rho^* \sim N^{-2}$.

\begin{figure}[H]
\centering
\includegraphics[trim={0.1cm 0.2cm 0.0cm 0.0cm},clip, width=0.48\textwidth]{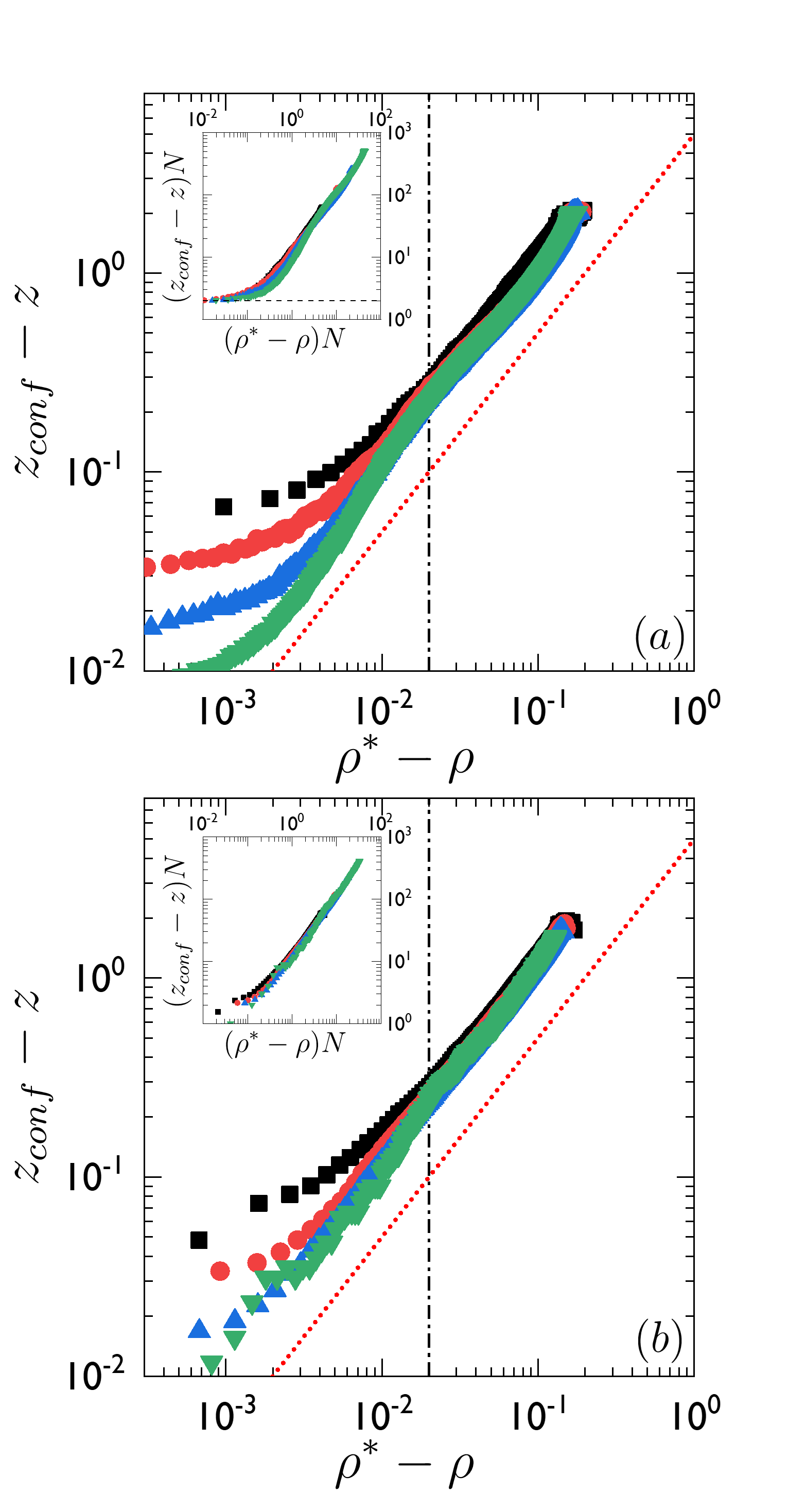}
\caption{The deviation in the contact number $z_{\rm conf} -z$ 
verus $\rho^*-\rho$, where $\rho^*$ is the 
packing fraction at which $z=z_{\rm conf}=6$, for (a) the SP and (b) DP 
models. For both the SP and DP models, we show a range of system sizes: $N = 32$ (squares), $64$ (circles), 
$128$ (upward triangles), and $256$ (downward triangles). The vertical 
dashed-dotted lines indicate the values of $\rho^*-\rho$ that correspond 
to the vertical dashed-dotted lines in Fig.~\ref{fig.4} (a). The insets 
show the same data as in the main panels except $z_{\rm conf}-z$ and 
$\rho^*-\rho$ are scaled by $N$. The horizontal dashed line in the inset to 
panel (a) is $(z_{\rm conf}-z)N = 2$.}  
\label{fig.6}
\end{figure}
\begin{table}[ht]
\caption{Parameters for the scaling forms for the excess contact number 
$\Delta z(\Delta \phi)$ and $\Delta z(\Delta \rho)$ 
(Eqs.~\ref{contact_number} and~\ref{zrho}) for the SP and DP models.}
\centering 
\begin{tabular}{c c c c c c c} 
\hline\hline 
Model & $z_J \pm 0.005 $ & $\phi_J,\rho_J \pm 0.005$ & $z_0^{\phi,\rho}$ & $\alpha_0^{\phi,\rho}$ & $z_1^{\phi,\rho}$ & $\alpha_1^{\phi,\rho}$\\ [0.5ex] 
\hline 
SP($\phi$) &  3.97  & 0.84 &  3.7 & 0.5&0&1.0 \\ 
SP($\rho$) &  3.97 &0.84  &2.6 & 0.5&4.0&1.0 \\
DP($\rho$)&  3.97& 0.835&3.3 & 0.5&7.1&1.0 \\ [1ex] 
\hline 
\end{tabular}
\label{table:nonlin1} 
\end{table}

\begin{table}[ht]
\caption{Parameters for the scaling forms for the shear modulus  
${\cal G}(\Delta \phi)$ and ${\cal G}(\Delta \rho)$ 
(Eqs.~\ref{shear_modulus} and~\ref{Grho}) for SP and DP models.} 
\centering 
\begin{tabular}{ c c c c c c} 
\hline\hline 
Model & $\phi_J,\rho_J  \pm 0.005$& ${\cal G}_0^{\phi,\rho}$ & $\beta_0^{\phi,\rho}$ & ${\cal G}_1^{\phi,\rho}$ & $\beta_1^{\phi,\rho}$\\ [0.5ex] 
\hline 
SP($\phi$) & 0.84&  0.42 & 0.5&0&1.0 \\ 
SP($\rho$)& 0.84  &0.38 & 0.5&0.35&1.0 \\
DP($\rho$) & 0.835&0.07& 0.5&1.7&1.0 \\[1ex] 
\hline 
\end{tabular}
\label{table:nonlin2} 
\end{table}

In Fig.~\ref{fig.5} (c) and (f), we show $\Delta z$ and ${\cal G}$
versus $\Delta \rho$ for the DP model. The data also show two regimes
in $\Delta \rho$. At small $\Delta \rho$, $\Delta z \sim \Delta \rho$
and ${\cal G} \sim \Delta \rho^0$. At intermediate and large $\Delta
\rho$, we fit $\Delta z$ and ${\cal G}$ to a sum of two power laws
with exponents $0.5$ and $1$. (See Tables~\ref{table:nonlin1}
and~\ref{table:nonlin2}.)  Note that the scaling of $\Delta z$ and
${\cal G}$ for the DP model is similar to that for the SP model
(Eqs.\ref{zrho} and \ref{Grho}), but the ratio of the coefficient for
the linear term in $\Delta \rho$ to that for the $\Delta \rho^{0.5}$
term is much larger than the corresponding ratio for the SP model. If
we use only a single power law exponent to fit the data for $\Delta z$
and ${\cal G}$ at large $\Delta \rho$, we find a best fit exponent of
$0.8$ for both $\Delta z$ and ${\cal G}$.  Winklemann, {\it et
  al.}~\cite{mobius_colloid} recently studied packings of bubbles and
showed that $z-z_J \sim (\rho - \rho_J)^{\alpha'} $ with an exponent
$\alpha' = 1.0$, which is larger than the value we find if we use a
single power-law to fit the data for $\Delta z$ for $\Delta \rho >
10^{-2}$.

In Fig.~\ref{fig.5} (a) and (b), the data for $\Delta z$ begin to
plateau at large $\Delta \phi$ or $\Delta \rho$, indicating that the
packings are reaching confluence with $z_{\rm conf} = 6$ at $\rho =
\rho^*$, where $\rho^* \rightarrow 1$ in the large system limit. To
understand the scaling of $z$ near confluence, we plot $z_{\rm
  conf}-z$ versus $\rho^*-\rho$. For the SP model, we use
Eq.~\ref{3body_rho} to determine $\rho$. We find that the data for
$z_{\rm conf}-z$ versus $\rho^*-\rho$ for different system sizes can
be collapsed onto a master curve when we scale $z_{\rm conf}-z$ and
$\rho^*-\rho$ by $N$ as shown in the inset to Fig.~\ref{fig.6}
(a). However, as $\rho \rightarrow \rho^*$, $(z_{\rm conf}-z)N$ does
not approach zero for the SP model. Instead, $(z_{\rm conf}-z)N
\rightarrow 2$. $(z_{\rm conf}-z)N$ approaches a finite value because
Eq.~\ref{3body_rho} does not account for all multiply-counted overlaps
between mutually overlapping pairs of disks (i.e. beyond three
mutually overlapping disks). For the DP model, $z_{\rm conf}-z$ verus
$\rho^* - \rho$ can also be collapsed onto a master curve when both
$z_{\rm conf}-z$ and $\rho^* - \rho$ are scaled by $N$
(Fig~\ref{fig.6} (b)).  Unlike the SP model, for the DP model we find that
$z_{\rm conf}-z$ scales as $\rho^* - \rho$ in the large system limit
over several orders of magnitude in $\rho^*-\rho$.

\begin{figure}[h]
\centering
\includegraphics[trim={0.0cm 0.0cm 0.0cm 0.0cm},clip, width=0.48\textwidth]{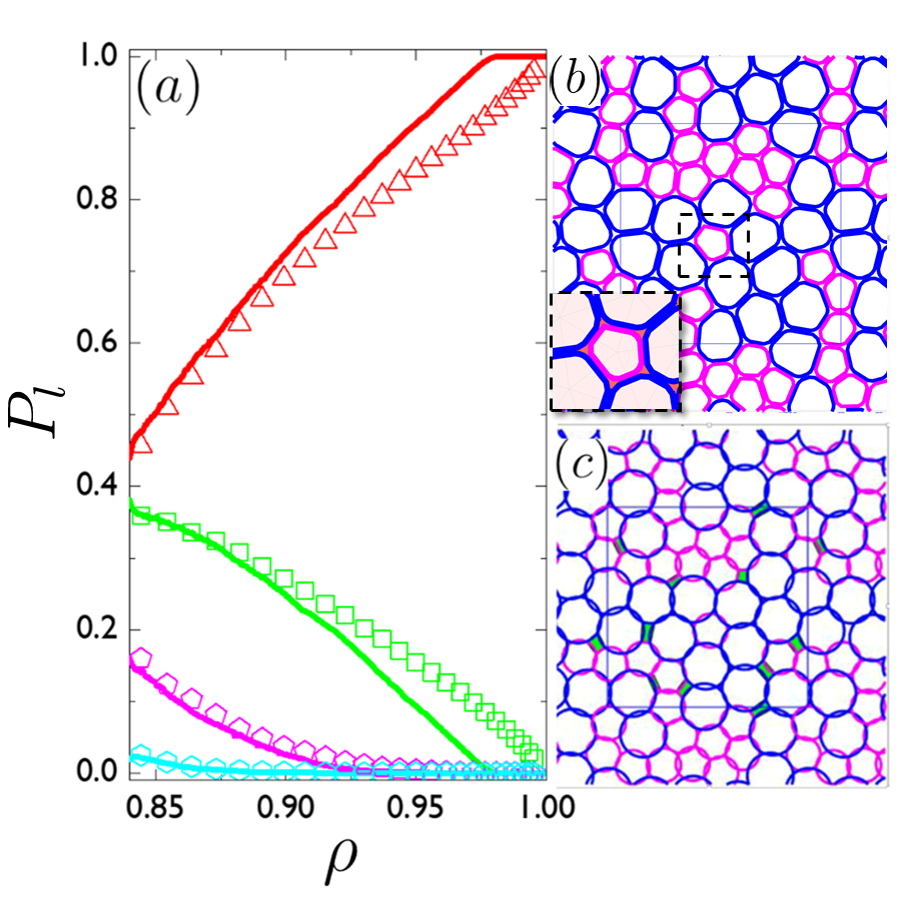}
\caption{(a) Probability $P_l$ to have a void with $l$ sides 
($l=3$ (red), $4$ (green), $5$ (pink), and $6$ (cyan))
for jammed packings as a function of $\rho$ for the SP (symbols) 
and DP models (lines). For both models, we studied ensembles of $500$ jammed packings 
of $N=64$ bidisperse particles. The void probability is normalized such that 
$\sum_{l} P_l(\rho) =1$. We also show snapshots of jammed packings 
using (b) the DP and (c) SP models at $\rho = 0.97$. The large (small) particles 
are outlined in blue (pink). $3$- and $4$-sided 
voids are shaded red and green, respectively. The square boxes with 
a solid outline indicate the main simulation cells. 
The inset in panel (b) is a close-up of the region within the small box
with a dashed outline.}
\label{fig.7}
\end{figure}

In addition to the contact number and shear modulus, we also studied
the variation of the structure of the void space of jammed packings as
a function of packing fraction $\rho$. To do this, for each jammed
packing, we identify all of the void space that is not occupied by
particles. We then determine the connected void regions (i.e. one can
reach any part of a connected void region from any point in the
region). The topology of each connected void can be characterized by
the number of edges, or the smallest number of particles that form a
loop on the perimeter of the void region (using the depth first search
algorithm).~\cite{dfs} In Fig.~\ref{fig.7} (a), we compare the
probability $P_l(\rho)$ to have a void with $l$ sides as a function of
$\rho$ for the SP and DP models. Near $\rho_J$, the SP and DP models
are identical and we find that the distributions $P_l(\rho_J)$ are the
same for the two models. In this regime, the probability of $3$- and
$4$-sided voids are similar ($\sim 0.4$), while the probabilities for
$5$- and $6$-sided voids are much smaller ($\sim 0.18$ and $\sim
0.02$). For $\rho \gtrsim 0.88$, $P_l(\rho)$ for the DP model begins
to deviate from that for the SP model.  Similar behavior was found for
the deviation in the contact number $\Delta z(\rho)$ for $\rho \gtrsim
0.88$. For the DP model, we find an increase in the probability of
$3$-sided voids over that for the SP model and a comparable decrease
in the probability of $4$-sided voids relative to the SP model. For
the DP model, we find that $P_3 =1$ for $\rho \gtrsim 0.97$. In
contrast, $P_3 =1$ only in the limit $\rho \rightarrow 1$ for the SP
model. Further, we find that the shape parameter ${\cal A} = {\cal
  A}_p = \pi/(4 \sqrt{3} - 2\pi) \approx 4.87$ of the $3$-sided voids
for the DP model is independent of $\rho$, indicating that the DP
model correctly captures the structure of the Plateau borders that
form as $\rho \rightarrow 1$.  However, the shape parameter of the
$3$-sided voids varies from ${\cal A} \approx {\cal A}_p$ to less than
$2$ as $\rho$ increases from $\rho_J$ to $1$ for the SP model, which
indicates that the void structure for the SP model differs
significantly from that for jammed packings of foams and emulsions
near confluence.

\begin{figure*}[t]
\centering
\includegraphics[trim={0.1cm 0.0cm 1.0cm 0.0cm},clip, width=\textwidth]{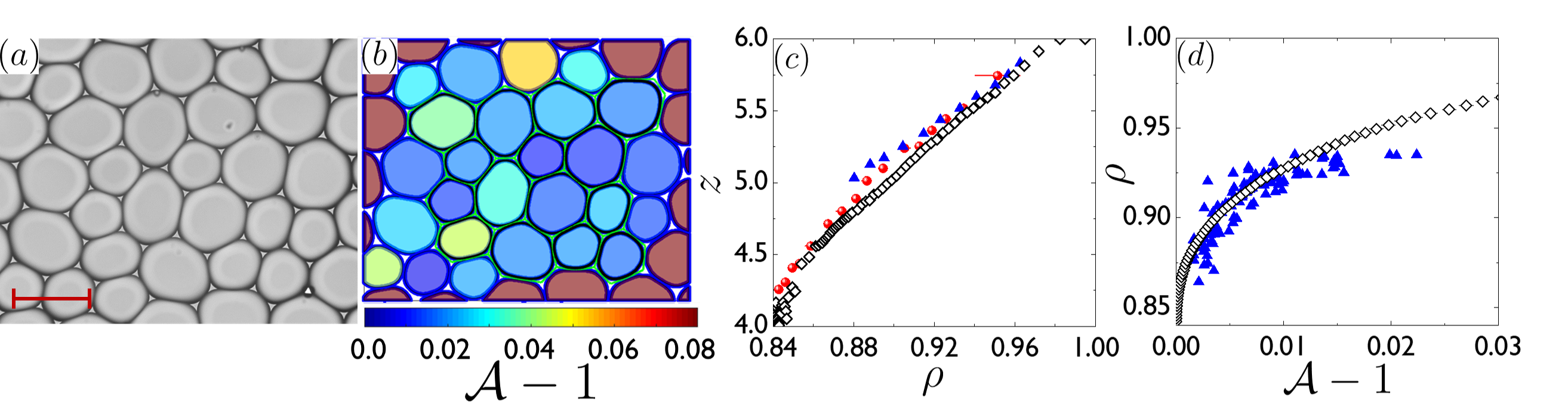}
\caption{(a) An example optical microscopy image of over-compressed quasi-2D oil
droplets in water at packing fraction
$\rho = 0.91\pm0.1$~\cite{desmond_sm}. The average droplet size is $D \approx 200\mu m$. The scale bar in (a) is $ 200\mu m$.  (b) Processed image in (a) from which we measured the 
shape parameter ${\cal A}-1$, contact number, and local packing fraction of 
each droplet using surface-voronoi tessellation (green lines). As ${\cal A}$ increases from $1$ to $1.08$ the color varies 
from dark blue to dark red. (c) Contact number $z$ plotted versus $\rho$ for the DP model 
(open diamonds) and experiments (filled triangles: emulsions~\cite{desmond_sm} and filled circles: foams~\cite{katgert_epl}). 
(d) $\rho$ versus ${\cal A}-1$ for the DP model (open diamonds). We also 
include a scatter plot of $\rho$ versus ${\cal A}-1$ for $\sim 150$ emulsion 
droplets (filled triangles: emulsions~\cite{desmond_sm}).}
\label{fig.8}
\end{figure*}

In Fig.~\ref{fig.8}, we show preliminary studies comparing the results
for jammed packings generated using the DP model to the results from
optical microscopy experiments of quasi-2D compressed emulsion
droplets~\cite{desmond_sm}.  We find that the contact number $z$ versus $\rho$ for the
DP model closely matches that from the experiments (Fig.~\ref{fig.8}
(c)). In addition, we show that the evolution of the shape factor of
the particles with packing fraction above jamming onset is similar for
the DP model and experiments (Fig.~\ref{fig.8} (d)). However, we
encourage additional experiments on compressed foams and emulsions to
be performed with packing fraction $\rho \gtrsim 0.95$ to determine whether
the DP model can recapitulate the structural and mechanical properties
of compressed emulsions near confluence.  For example, new
experimental studies can test the prediction for the DP model that
$z_{\rm conf}-z$ scales linearly with $1 - \rho$ near $\rho = 1$.

\section{Discussion and Conclusions}
\label{conclusions}

In this article, we investigated the structural and mechanical properties of jammed packings undergoing isotropic compression in 2D using the soft particle (SP) model~\cite{durian_prl, makse_prl} and the new deformable particle model~\cite{arman_prl} that we developed for bubbles and emulsions. The SP model has been widely used to characterize the structural, mechanical, and rheological properties of jammed particulate systems including granular materials~\cite{corey_pre}, dense colloidal suspensions~\cite{kchen_prl}, foams~\cite{durian_prl}, and emulsions~\cite{makse_pre}. The key difference between the
two models is that in the SP model, the particles decrease in area as
the system is compressed above jamming onset, while the DP model
conserves particle area during compression. Studies that have employed
the SP model typically characterize the properties of jammed packings
as a function of the reduced number density $\phi$~\cite{corey_prl,
  liu_rev,vanhecke_rev}, rather than using the true packing fraction
$\rho$.  In this work, we provide direct comparisons of the structural
and mechanical properties of packings generated using the SP and DP
models as a function of the packing fraction $\rho$.

First, we showed explicitly that the SP and DP models give the same
results near jamming onset, where the disks are undeformed.  In
particular, we showed that the probability distribution $P(\rho_J)$ of
jamming onsets $\rho_J \approx \phi_J$ are nearly identical for the SP
and DP models with $\rho_J \approx 0.842$ in the large system
limit. (For the detailed discussion of this point, see the Appendix.)
In addition, we find similar scaling behavior for the excess contact
number $z-z_J$ and shear modulus ${\cal G}$ versus $\rho-\rho_J$. Near
jamming onset, for both the SP and DP models, $z-z_J \sim
(\rho-\rho_J)^{1.0}$ and ${\cal G} \sim (\rho-\rho_J)^{0}$ for small
systems.~\cite{goodrich_prl,goodrich_pre} This scaling behavior occurs
for $\rho -\rho_J < \Delta \rho^*$, where $\Delta \rho^* \sim
N^{-2}$. In the large system limit, $z-z_J \sim {\cal G} \sim
(\rho-\rho_J)^{0.5}$ near jamming onset for both the SP and DP models.

For packings that are compressed above jamming onset, we determined
the relation between the packing fraction $\rho$ and reduced number
density $\phi$ for the SP model. Using this relation, we showed that
for the SP model $\Delta z(\Delta \rho) \sim {\cal G}(\Delta \rho)$
can be represented as a sum of power-laws in $\Delta \rho$ (not a single
power-law), with $\Delta \rho^{0.5}$ and $\Delta \rho^{1.0}$ as the
lowest order terms.  The scaling of $\Delta z$ and ${\cal G}$ is
similar for the DP model, but the ratio of the coefficient for the
linear term in $\Delta \rho$ to that for the $\Delta \rho^{0.5}$ term
is much larger than the corresponding ratio for the SP model.  As a result, we find
that $z(\rho)$ and ${\cal G}(\rho)$ are larger for the DP model
compared to the SP model for $\rho \gtrsim 0.88$.  

In addition, we characterized the void space in jammed
packings~\cite{bare_arxiv, torquado_pre} as a function of packing
fraction for both models. We showed that the DP model can recapitulate
the formation of Plateau borders as the deformable particles become
tightly packed.  The probability to obtain $3$-sided voids becomes
unity for $\rho \gtrsim 0.97$ and the shape factor of the $3$-sided
voids (${\cal A} \approx {\cal A}_p$) is independent of $\rho$. In
contrast, for the SP model, the probability of $3$-sided voids only
becomes unity in the limit $\rho \rightarrow 1$ and ${\cal A} < {\cal
  A}_p$ over the full range of packing fraction. Thus, the SP model
does not capture the topological features of the void space of
packings of compressible bubbles near confluence. We believe that the
results from this article will inspire additional experimental and
theoretical studies of the collective behavior of droplets and bubbles
in emulsions and foams. For example, the DP model can be extended to
include attractive interactions to investigate the mechanical response
of attractive emulsions in both 2D and 3D~\cite{jorjadze_pnas}. In
addition, active forces can be added to each circulo-line in the DP
model to simulate collective motion, such as swarming and migration,
in cell aggregates as well as living tissues.

\section{Appendix}

\begin{figure*}
 \centering
 \includegraphics[trim={0.1cm 0cm 0.0cm 0.0cm},clip, width=\textwidth]{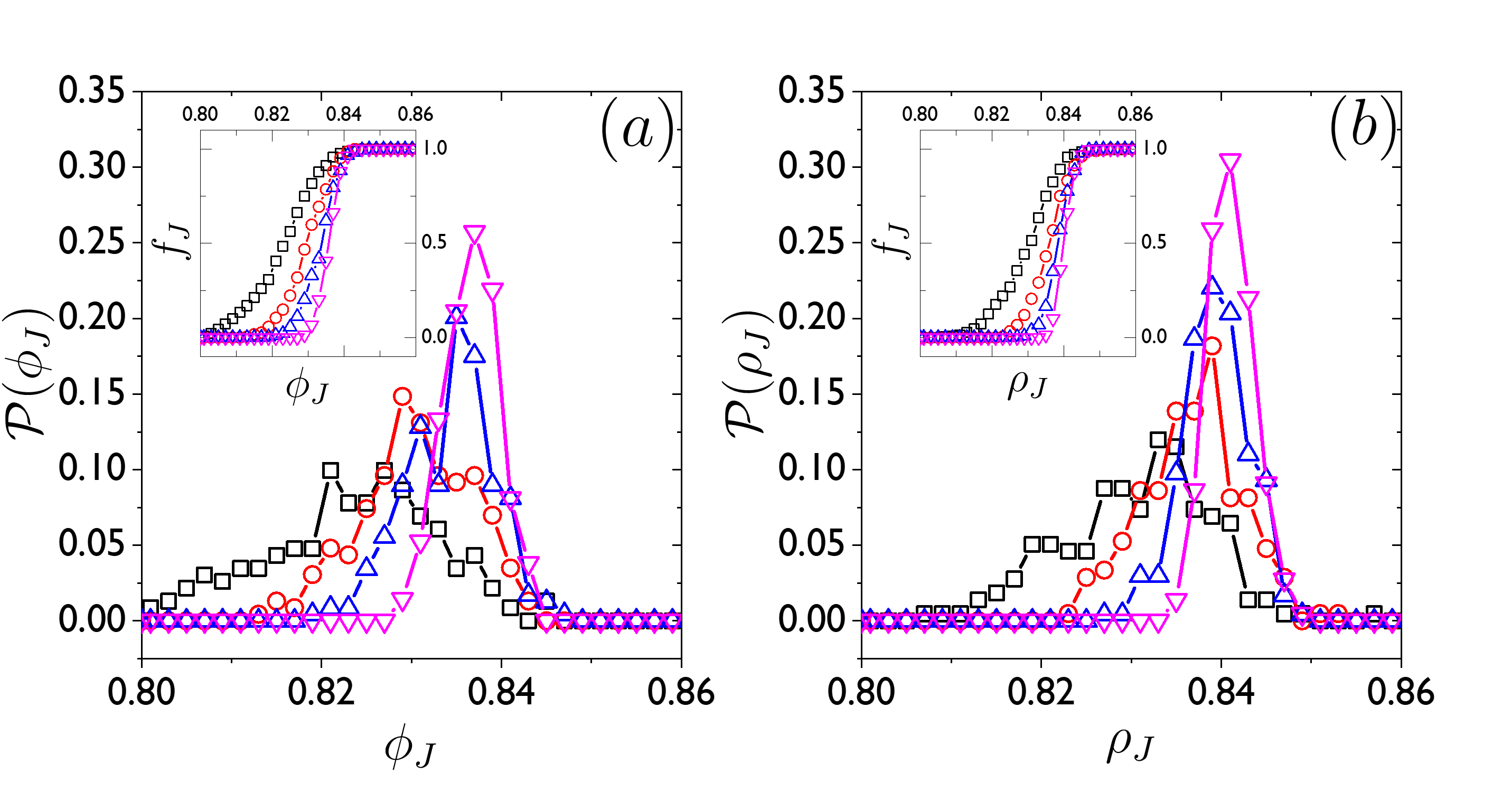}
\caption{(a) The probability distribution $P(\phi_J)$ to have reduced number 
density $\phi_J$ at jamming onset for the SP model and (b) the probability 
distribution $P(\rho_J)$ to have packing fraction $\rho_J$ at 
jamming onset for the DP model for several systems sizes: $N= 32$ (squares), 
$64$ (circles), $128$ (upward triangles), and $256$ (downward triangles). 
The insets in both panels give the fraction $f_J$ of the packings (out of 
$500$) that are jammed at or below a given $\phi_J$ (or $\rho_J$) for the 
SP and DP models.}
 \label{app.1}
\end{figure*}

In this Appendix, we provide additional calculations to support the
conclusions described in the main text.  In Fig.~\ref{app.1}, we show
the distribution of the reduced number densities $P(\phi_J)$ for the SP model
(panel (a)) and the distribution of packing fractions $P(\rho_J)$ for
the DP model (panel (b)) at jamming onset using the isotropic
compression protocol discussed in Sec.~\ref{protocol}. (We also
include the fraction $f_J$ of packings with $\phi_J$ (or $\rho_J$) at or below
a given value in the insets to Fig.~\ref{app.1} (a) and (b).) The
distributions for the SP and DP models were obtained using the same
initial random particle positions. At jamming onset $\phi_J = \rho_J$, and thus
the distributions $P(\phi_J)$ and $P(\rho_J)$ for the SP and DP models
are nearly identical. Similar to previous studies, we also find that the
root-mean-square deviation in the packing fraction at jamming 
onset scales as $\Delta \phi_J \sim N^{-\theta}$,
with $\theta \approx 0.55$, and $\phi_J \rightarrow 0.842$ in the
large system limit.

\begin{figure*}
 \centering
 \includegraphics[trim={0.1cm 0.2cm 0.0cm 0.0cm},clip, width=\textwidth]{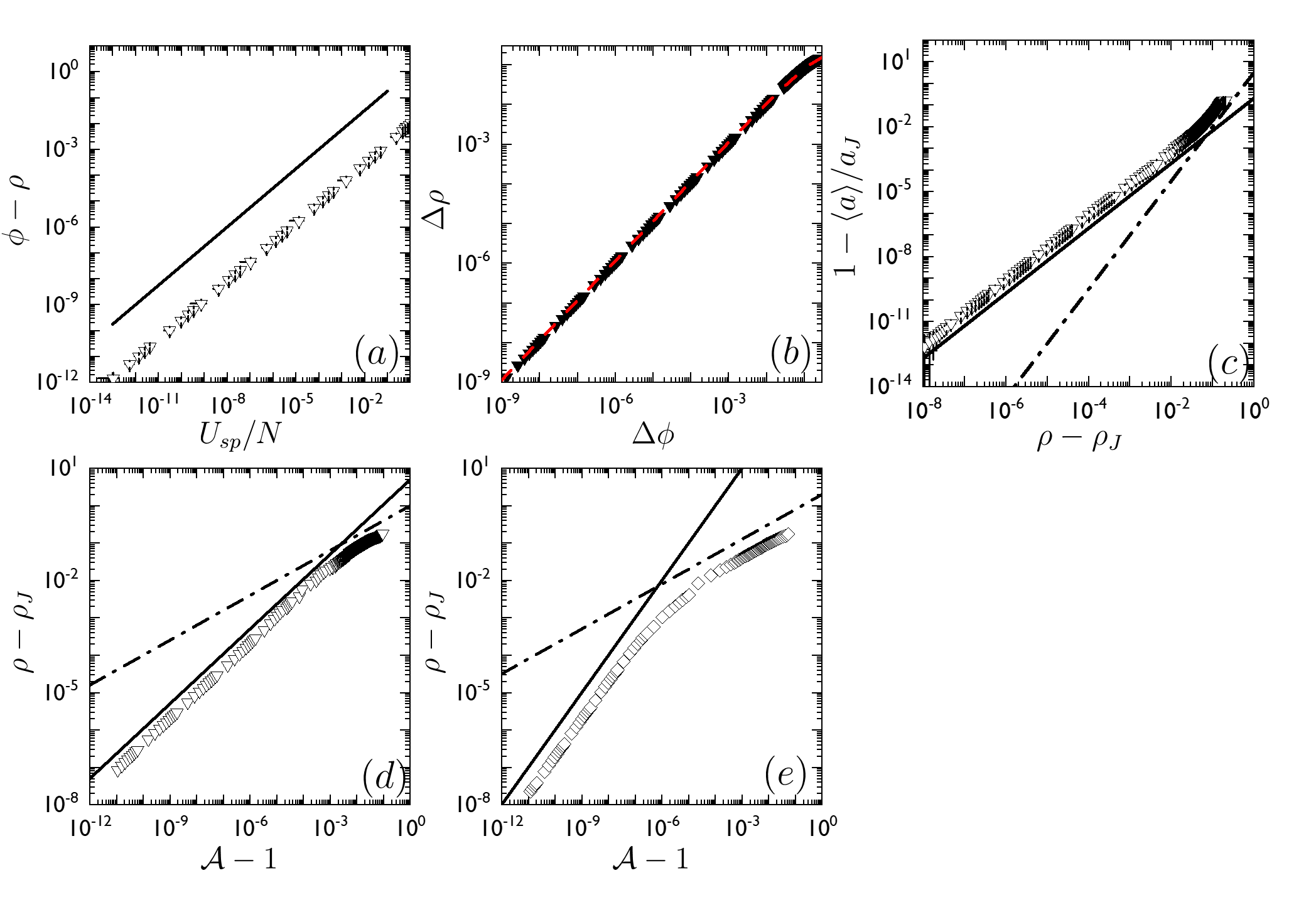}
 \caption{(a) The difference between the reduced number density, $\phi$, and 
true packing fraction $\rho$ versus the total potential energy per particle, 
$U_{SP}/N$, averaged over $500$ jammed packings of $N=32$ disks generated 
using SP model. The solid line has slope equal to $0.75$. (b) $\Delta \rho =
\rho-\rho_J$ plotted versus $\phi -\phi_J$ for the SP model. The dashed line 
represents Eq.\ref{exact} with $C = 1.21$.  (c) The relative 
deviation in the particle area from that at jamming 
onset, $1-\langle a \rangle/a_J$ plotted versus $\rho-\rho_J$ for the SP 
model. 
The solid and dashed-dotted lines have slopes 
equal to $1.5$ and $2.5$, respectively. (d) $\rho -\rho_J$ plotted versus the 
shape parameter, ${\cal A} - 1$, for the (d) SP and (e) DP models. 
The solid and dashed-dotted lines in panel (d) have slopes 
equal to $0.67$ and $0.3$, respectively. 
The solid and dashed-dotted lines in panel (e) have slopes equal to $1$
and $0.3$, respectively.}
 \label{app.2}
\end{figure*}

In Fig.~\ref{app.2}, we show a series of scaling laws for several
physical quantities and compare the behavior for packings generated
using the SP and DP models. In Fig.~\ref{app.2} (a), we show that
$\rho - \phi \sim (U_{SP}/N)^{\kappa}$, where $\kappa \approx 0.75$,
for the SP model.  The scaling exponent $\kappa$ can be obtained by
assuming $\phi - \rho \propto \Delta\phi^{1.5}$ from Eq.~\ref{exact}
and using $\Delta\phi \sim (U_{SP}/N)^{0.5}$ from Eq.~\ref{tote2s}.

In Fig.~\ref{app.2} (b), we show $\Delta \rho = \rho -\rho_J$ versus
$\Delta\phi = \phi-\phi_J$ for packings generated using the SP
model. The dashed line obeys Eq.~\ref{exact} with $C \approx 1.2$. As
discussed in Sec.~\ref{results}, the relation between $\Delta\rho$ and
$\Delta \phi$ only holds when multiply-counted overlaps (i.e. beyond
pairwise overlaps) are absent. Combining the results from
Fig.~\ref{app.2} (b) and~\ref{fig.3} (a), we see that Eq.~\ref{exact}
holds over nearly $10$ orders of magnitude in $\Delta \phi$. In
Fig.~\ref{app.2} (c), we show the data for the relative deviation in
the area of the particles $1-\langle a\rangle/a_J$ versus
$\rho-\rho_J$ from Fig.~\ref{fig.3} (b) on logarithmic scales for the
SP model. In contrast to the DP model, the SP model does not conserve
particle area when ovecompressed above jamming onset.  We find two
power-law scaling regimes. $1-\langle a\rangle/a_J \sim \Delta
\rho^{\zeta}$ with $\zeta \approx 1.5$ and $2.5$ at small and
large $\Delta \rho$, respectively.

In Figs.~\ref{app.2} (c) and (d), we characterize the change in particle 
shape as the packings are compressed above jamming onset for the SP and 
DP models. We show that $\Delta \rho \sim ({\cal A-}1)^{\omega}$, but the 
scaling exponent takes on different values in the for small and large values 
of ${\cal A}-1$. At small ${\cal A}-1$, $\omega \approx 0.66$ and $\approx 
1.0$ for the SP and DP models, respectively. At large ${\cal A}-1$, $\omega \approx 0.3$ for both the SP and DP models.  

\section{Acknowledgments}
\label{acknowledgments}
We acknowledge support from NSF Grants No. PHY-1522467 (A.B., A.S., and 
C.S. O'Hern),
No. CBET-1336401 (J.L., C.S. Orellana, and E.R.W),  No. CBET-1804186 (E.R.W),
No. CMMI-1462439 (C.S. O'Hern) and No. CMMI-1463455 (M.D.S.), the President's 
International Fellowship Initiative (PIFI) and Hundred Talents Program of 
the Chinese Academy of Sciences (A.B. and F.Y.), and the Raymond 
and Beverly Sackler Institute for Biological, Physical, and Engineering 
Sciences (A.B.). This work was also supported by the High Performance Computing
facilities operated by Yale's Center for Research Computing. 

\section*{Conflicts of interest}
There are no conflicts to declare.



\balance

\renewcommand\refname{References}

\clearpage
\scriptsize{
\providecommand*{\mcitethebibliography}{\thebibliography}
\csname @ifundefined\endcsname{endmcitethebibliography}
{\let\endmcitethebibliography\endthebibliography}{}

\end{document}